\documentclass[a4paper,UKenglish,cleveref, autoref]{lipics-v2019}

\title{Optimal Time and Space Construction of Suffix Arrays and LCP Arrays for Integer Alphabets}

\titlerunning{Optimal Time and Space Construction of Suffix Arrays and LCP Arrays}

\author{Keisuke Goto}{Fujitsu Laboratories Ltd.\\{Kawasaki, Japan}}{goto.keisuke@fujitsu.com}{https://orcid.org/0000-0001-6964-6182}{}

\authorrunning{K. Goto}

\Copyright{Keisuke Goto}

\ccsdesc[100]{Theory of computation, Pattern matching}

\keywords{Suffix Array, Longest Common Prefix Array, In-Place Algorithm}

\acknowledgements{
We wish to thank Takashi Kato, Shunsuke Inenaga, Hideo Bannai, Dominik K\"oppl, and anonymous reviewers for their many valuable suggestions on improving the quality of this paper, and especially, we also wish to thank an anonymous reviewer for giving us a simpler algorithm for computing LCP arrays as described in Section~\ref{sec:lcp}.
}

\nolinenumbers 

\hideLIPIcs  

\EventEditors{John Q. Open and Joan R. Access}
\EventNoEds{2}
\EventLongTitle{42nd Conference on Very Important Topics (CVIT 2016)}
\EventShortTitle{CVIT 2016}
\EventAcronym{CVIT}
\EventYear{2016}
\EventDate{December 24--27, 2016}
\EventLocation{Little Whinging, United Kingdom}
\EventLogo{}
\SeriesVolume{42}
\ArticleNo{23}

\usepackage{booktabs}
\usepackage{hyperref}
\usepackage{appendix}
\usepackage{cite}
\usepackage[whole]{bxcjkjatype}
\usepackage{xcolor}
 \usepackage[final,mode=multiuser]{fixme}
\usepackage{graphicx}
\usepackage{amsthm}
\usepackage{xspace}
\usepackage{comment}

\fxsetup{theme=color}
\fxuseenvlayout{colorsig}
\fxusetargetlayout{color}
\FXRegisterAuthor{kg}{akg}{\color{red}KG}
\definecolor{kgnote*}{rgb}{1.0, 0, 0}
\definecolor{fxtarget}{rgb}{1.0, 0, 0}
\colorlet{kgnotebg*}{red}
\colorlet{kgnotebg}{red}
\fxsetface{env}{}
\fxsetface{inline}{\bfseries}
\fxsetface{margin}{\tiny}

\newcommand{\attention}[1]{{\color{red}#1}}

\newcounter{tcounter}
\newcounter{counterMakeSALMS}
\newcounter{counterMakeSALMSynot}
\newcounter{counterMakeYLMSy}
\newcounter{counterMakeYLMSx}
\newcounter{counterMakeSALMSxnot}
\newcounter{counterMakeLFLMS}
\newcounter{counterMakeSALxnot}
\newcounter{counterMakeSALx}
\newcounter{counterMakeSAL}

\stepcounter{tcounter}
\stepcounter{tcounter}
\stepcounter{tcounter}
\stepcounter{tcounter}
\stepcounter{tcounter}
\stepcounter{tcounter}
\stepcounter{tcounter}
\stepcounter{tcounter}
\newcommand{\transition}[1]{ \textbf{Transition #1:\xspace} }
\newcommand{\makeSALMS}{\transition{\arabic{counterMakeSALMS}}}
\newcommand{\makeSALMSynot}{\transition{\arabic{counterMakeSALMSynot}}}
\newcommand{\makeYLMSy}{\transition{\arabic{counterMakeYLMSy}}}
\newcommand{\makeYLMSx}{\transition{\arabic{counterMakeYLMSx}}}
\newcommand{\makeSALMSxnot}{\transition{\arabic{counterMakeSALMSxnot}}}
\newcommand{\makeLFLMSx}{\transition{\arabic{counterMakeLFLMS}}}
\newcommand{\makeSALxnot}{\transition{\arabic{counterMakeSALxnot}}}
\newcommand{\makeSALx}{\transition{\arabic{counterMakeSALx}}}
\newcommand{\makeSAL}{\transition{\arabic{counterMakeSAL}}}

\renewcommand{\dots}{\ldots}

\newcommand{\ii}{i}
\newcommand{\jj}{j}
\newcommand{\kk}{k}
\newcommand{\vt}{t}

\newcommand{\N}{N}
\newcommand{\mm}{m}

\newcommand{\T}{{\bf T}}
\newcommand{\SA}{{\bf SA}}

\newcommand{\LCP}{{\bf LCP}}

\newcommand{\psia}{{\bf \Psi}}

\newcommand{\A}{{\bf A}}
\newcommand{\B}{{\bf B}}

\newcommand{\X}{{\bf X}}
\newcommand{\Y}{{\bf Y}}
\newcommand{\Z}{{\bf Z}}
\newcommand{\lf}{{\bf LE}}
\newcommand{\rf}{{\bf RE}}
\newcommand{\typelf}{{\bf type}}

\newcommand{\suf}[1]{\T_{#1}}
\newcommand{\sub}[1]{\T^\prime_{#1}}
\newcommand{\type}{{\bf type}}
\newcommand{\num}[1]{N_{#1}}

\newcommand{\headc}[1]{t_{#1}}
\newcommand{\rank}{\mathit{rank}}

\newcommand{\setsuf}[1]{\mathit{suf}(\mathit{#1})}
\newcommand{\setall}{\setsuf{all}}

\newcommand{\setLML}{\setsuf{LML}}

\newcommand{\setLMS}{\setsuf{LMS}}
\newcommand{\setxLMS}{\setsuf{LMSx}}
\newcommand{\setxLMSnot}{\setsuf{\overline{LMSx}}}
\newcommand{\setyLMS}{\setsuf{LMSy}}
\newcommand{\setyLMSnot}{\setsuf{\overline{LMSy}}}

\newcommand{\setL}{\setsuf{L}}
\newcommand{\setxL}{\setsuf{Lx}}
\newcommand{\setxLnot}{\setsuf{\overline{Lx}}}
\newcommand{\setyL}{\setsuf{Ly}}
\newcommand{\setyLnot}{\setsuf{\overline{Ly}}}

\newcommand{\setS}{\setsuf{S}}
\newcommand{\setxS}{\setsuf{Sx}}
\newcommand{\setxSnot}{\setsuf{\overline{Sx}}}

\newcommand{\CL}{\mathit{CL}}
\newcommand{\C}{\mathit{C}}

\newcommand{\floor}[1]{\lfloor #1 \rfloor}
\newcommand{\ceil}[1]{\lceil #1 \rceil}
\newcommand{\full}{{\bf B}}

\renewcommand{\attention}[1]{}

\begin{document}

\maketitle
\begin{abstract}
  Suffix arrays and LCP arrays are one of the most fundamental data structures widely used for various kinds of string processing.
  We consider two problems for a read-only string of length $N$ over an integer alphabet $[1, \dots, \sigma]$ for $1 \leq \sigma \leq N$, the string contains $\sigma$ distinct characters, the construction of the suffix array, and a simultaneous construction of both the suffix array and LCP array.
  For the word RAM model, we propose algorithms to solve both of the problems in $O(N)$ time by using $O(1)$ extra words, which are optimal in time and space.
  Extra words means the required space except for the space of the input string and output suffix array and LCP array.
  Our contribution improves the previous most efficient algorithms, $O(N)$ time using $\sigma+O(1)$ extra words by [Nong, TOIS 2013] and $O(N \log N)$ time using $O(1)$ extra words by [Franceschini and Muthukrishnan, ICALP 2007], for constructing suffix arrays, and it improves the previous most efficient solution that runs in $O(N)$ time using $\sigma + O(1)$ extra words for constructing both suffix arrays and LCP arrays through a combination of [Nong, TOIS 2013] and [Manzini, SWAT 2004].
\end{abstract}

\section{Introduction}
Suffix arrays~\cite{ManberM93} are data structures that store all suffix positions of a given string sorted in lexicographical order according to their corresponding suffixes.
They were proposed as a space efficient alternative to suffix trees, which are one of the most fundamental and powerful tools used for various kinds of string processing.
LCP arrays~\cite{ManberM93} are auxiliary data structures that store the lengths of the longest common prefixes between adjacent suffixes stored in suffix arrays.
Suffix arrays with LCP arrays are sometimes called \textit{enhanced suffix arrays}~\cite{AbouelhodaKO02}, and they can simulate various operations of suffix trees.
Suffix arrays or enhanced suffix arrays can be used for efficiently solving problems in various research areas, such as pattern matching~\cite{ManberM93,Navarro01}, genome analysis~\cite{AbouelhodaKO02,LiS01}, text compression~\cite{GotoB14, CrochemoreI08,Burrows94ablock-sorting}, and data mining~\cite{GotoBIT13,FischerHK05}.
In these applications, one of the main computational bottlenecks is the time and space needed to construct suffix arrays and LCP arrays.

In this paper, we consider two problems that are for a given read-only string: constructing suffix arrays and constructing both suffix arrays and LCP arrays.
For both problems, we propose optimal time and space algorithms.
We assume that an input string of length $\N$ is read only, consists of an integer alphabet $[1, \dots, \sigma]$ for $1 \leq \sigma \leq \N$, and contains $\sigma$ distinct characters~\footnote{As we will describe later, this is a slightly stronger assumption than commonly used in previous research.}.
We assume that the word RAM model with a word size of $w=\ceil{\log N}$ bits and that basic arithmetic and bit operations on constant number of words take constant time.
We say that an algorithm runs \textit{in-place} and runs in optimal space if the algorithm requires constant extra words, which is the space except for the input string, output suffix array, and LCP array.

\textbf{Suffix array construction.}
The first linear time (optimal time) algorithms for constructing suffix arrays for a string over an integer alphabet $[1, \dots, \sigma]$ for $1 \leq \sigma \leq \N$ were proposed at the same time by several authors~\cite{KoA05, KarkkainenSB06, KimSPP05}, and they require at least $\N$ extra words.
Nong~\cite{Nong13} proposed a linear time and space efficient algorithm that requires $\sigma + O(1)$ extra words, but it still requires about $N$ extra words in the worst case since $\sigma$ can be~$N$.
An in-place algorithm that runs in $O(N \log N)$ time was proposed by Franceschini and Muthukrishnan~\cite{FranceschiniM07}~\footnote{They assume that an input string is over a general alphabet, i.e., only comparison of any two characters is allowed, which can be done in $O(1)$ time. Their time and space complexities are optimal for general alphabets but not for integer alphabets.}.
It has been an open problem whether there exists a suffix array construction algorithm that runs in linear time and in-place.

We propose an in-place linear time algorithm for a string over an integer alphabet $[1, \dots, \sigma]$ that consists of $\sigma$ distinct characters.
Our algorithm is based on the induced sorting framework~\cite{KoA05,NongZC11,Nong13,FranceschiniM07}, which splits all suffixes into L- and S-suffixes, sorts either of which first, and then sorts the other.
The induced sorting framework uses two arrays: (1) a bit array of $\N$ bits to store each type of suffix, and (2) an integer array of $\sigma$ words, for each character $\vt$, to store a pointer to the next insertion position of a suffix starting with $\vt$ in the suffix array, so this framework requires $\sigma + \N / \log \N + O(1)$ extra words in a naive way.
Our algorithm runs in almost the same way as the previous ones~\cite{KoA05,NongZC11,Nong13,FranceschiniM07}, but it stores these two arrays in the space of the output suffix array.
Therefore, our algorithm runs in linear time and in-place.
As a minor contribution, we also propose a simple space saving technique for the induced sorting framework.
The framework has to store the beginning and ending positions of sub-arrays in recursive steps, which requires $O(\log \N)$ words in total in a naive way.
Franceschini and Muthukrishnan~\cite{FranceschiniM07} proposed a method for storing them in-place and obtained each value in $O(\log \N)$ time.
We propose a simpler one for storing them in-place and obtain each value in $O(1)$ time, and we describe it in Appendix~\ref{sec:recursion}.

Our assumption is slightly stronger than those of previous research in that all characters of an alphabet must appear in the input string~\footnote{The same problem setting also appeared in~\cite{BarbayCGNN14}.}.
However, if an input string can be writable not read-only, our algorithm still runs in linear time and in-place also for the same problem setting to previous research which an input string is over an integer $[1, \dots, \sigma]$ and some characters may not appear in the string.
Because we can transform the string to a string over an integer alphabet $[1, \dots, \sigma^\prime]$ that consists of $\sigma^\prime \leq \sigma$ distinct characters by the counting sort~\cite{Knuth98aCountingSort} that uses the space of output suffix arrays before our algorithm runs.

\textbf{Recent and independent works for suffix array construction.}
Recently and independently, some in-place suffix array construction algorithms were proposed.

Li et al.~\cite{LiLH18} proposed an in-place linear time algorithm for a read-only string over an integer alphabet $[1, \dots, O(\N)]$ whose assumption is more general, and the result is stronger than ours.
Though Li et al.'s algorithm is also based on the induced sorting framework, the details are different from ours.
Both their and our algorithm look simple according to the framework, but this simplicity comes from using complex data structures and algorithms as tools. 
Li et al.'s algorithm uses two complex tools, in-place stable merge sort algorithm~\cite{Chen03} and succinct data structures supporting select queries in constant time~\cite{Jacobson89} which is used for storing pointers in compressed space.
On the other hand, our algorithm uses only the former one and store pointers in a normal array.
Using complex tools tend to increase the runtime in practice, e.g., according to~\cite{DelprattRR06}, select query times on a succinct data structure are several time slower than normal array accesses.
In this perspective, our algorithm is simpler than theirs, and our work may contribute to develop practically faster in-place linear time suffix array construction algorithms in future. 

Prezza~\cite{Prezza18} studied a similar problem that, for a writable input string, sorts suffixes of a size-$b$ subset instead of all suffixes and constructs a \textit{sparse} suffix array and \textit{sparse} LCP array.
His algorithm is based on a longest common extension data structure that, for two given positions $\ii$ and $\jj$, efficiently computes the length of the longest common prefix between two suffixes starting at $\ii$ and $\jj$, and it runs in $O(\N + b \log^2 \N)$ expected time and in-place.

\textbf{Suffix array and LCP array construction.}
Most previous research focused on a setting that computes LCP arrays from a given string and its suffix array.
Kasai et al.~\cite{KasaiLAAP01} proposed the first linear time (optimal time) algorithm that computes the inverse suffix array and then uses it and computes the LCP array.
Since it stores the inverse suffix array in extra space, it requires $\N+O(1)$ extra words.
Manzini~\cite{Manzini04} proposed a more space efficient linear time algorithm.
The algorithm constructs the $\psi$ array, which is similar to the inverse suffix array, in the output space of the LCP array and then rewrites it to the LCP array.
The rewriting process runs in-place, but constructing the $\psi$ array requires $\sigma + O(1)$ extra words, so the algorithm runs in linear time using $\sigma + O(1)$ words in total.
Suffix arrays and LCP arrays can be computed in the same time and space by computing the suffix array with Nong's algorithm~\cite{Nong13} and then by computing the LCP array with Manzini's algorithm~\cite{Manzini04}.
The problem for constant alphabets with $\sigma \in O(1)$ has been studied in~\cite{LouzaGT17,Fischer11}, and the algorithms in~\cite{LouzaGT17,Fischer11} are very competitive in practice for constant alphabets.

Our proposed linear time in-place algorithm constructs the suffix array and LCP array on the basis of a simple but non-trivial strategy.
First, we construct the $\psi$ array by using the space of the suffix array and LCP array and store it in the space of the LCP array.
Then, we construct the suffix array in-place by using our linear time in-place algorithm, and we rewrite the $\psi$ array to the LCP array as in Manzini's algorithm.
Thus, we finally obtain both the suffix array and LCP array in linear time and in-place.

\textbf{Organization.}
This paper is organized as follows.
In Section~\ref{sec:preliminaries}, we introduce notations and definitions.
In Section~\ref{sec:previous}, we explain the induced sorting framework on which our algorithms are based.
In Section~\ref{sec:new_algo} and Section~\ref{sec:lcp}, we propose optimal time and space algorithms for constructing suffix arrays and both suffix arrays and LCP arrays, respectively.

\section{Preliminaries}
\label{sec:preliminaries}
\begin{figure}[tb]
	\centering
	\includegraphics[width=0.8\linewidth]{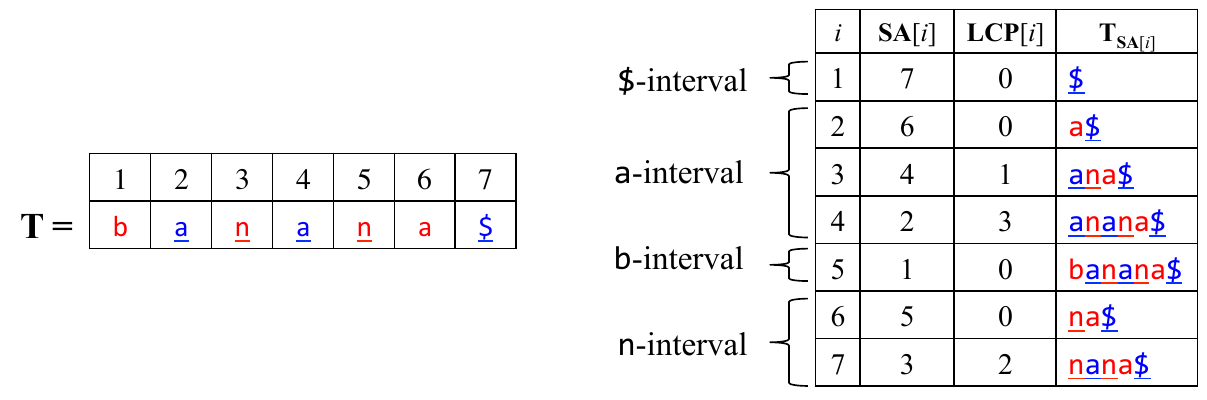}
	\caption{
		Suffix array and LCP array of $\T=\mathtt{banana\$}$.
		$\T[\ii]$ is colored red if $\suf{\ii}$ is an L-suffix, and blue otherwise.
		Moreover, $\T[\ii]$ is underlined if $\suf{\ii}$ is an LML- or LMS-suffix.
	}
	\label{fig:example_of_SA}
\end{figure}

Let $\Sigma$ be an integer alphabet, the elements of which are $[1, \dots, \sigma]$ for an integer $\sigma \geq 1$.
An element of $\Sigma^*$ is called a \textit{string}.
The length of a string $\T$ is denoted by $|\T|$.
The empty string $\epsilon$ is a string of length 0.
For a string $\T=xyz$, $x$, $y$ and $z$ are called a \textit{prefix}, \textit{substring}, and \textit{suffix} of $\T$, respectively.
For a string $\T$ of length $\N$, the $\ii$-th character of $\T$ is denoted by $\T[\ii]$ for $1 \leq \ii \leq \N$, and the substring of $\T$ that begins at position $\ii$ and that ends at position $\jj$ is denoted by $\T[\ii\dots \jj]$ for $1 \leq \ii \leq \jj \leq N$.
For convenience, we assume that $\T[\N]=\$$, where $\$$ is a special character lexicographically smaller than any characters in the string $\T[1 \dots \N-1]$.
In this paper, we also assume that $\sigma \leq \N$ and that $\T$ contains $\sigma$ distinct characters.

A suffix starting at a position $i$ and its first character are denoted by $\suf{i}$ and $\headc{i}$, respectively, and the position $i$ is also called a \textit{pointer} to $\suf{i}$.
If no confusion occurs, we sometimes use $\suf{i}$ as a pointer $i$.
For $1 \leq i \leq N$, $\suf{i}$ is called a \textit{small type suffix} (S-suffix) if $i=N$ or $\suf{i}$ is lexicographically smaller than $\suf{i+1}$, and it is called a \textit{large type suffix} (L-suffix) otherwise.
An S-suffix/L-suffix $\suf{i}$ is also called a \textit{leftmost-S-suffix}/\textit{leftmost-L-suffix} (LMS-suffix/LML-suffix) if $i > 1$ and $\suf{i-1}$ is an L-suffix/S-suffix.
For $i<N$, $\sub{i}$ denotes the substring $\T[i \dots j]$, where $j > i$ is the leftmost position such that  $\suf{j}$ is an LMS-suffix.
Each type of $\sub{i}$ is equal to that of $\suf{i}$, and $\sub{i}$ is referred to as an L-, S-, LML-, or LMS-substring according to its type.
The important property is that, for $1\leq i<N$,  $\suf{i}$ is an L-suffix if $\headc{i}>\headc{i+1}$, or $\headc{i}=\headc{i+1}$ and $\suf{i+1}$ is an L-suffix, and $\suf{i}$ is an S-suffix otherwise.
From this property, each type of suffix can be obtained in $O(N)$ time with a right-to-left scan on $\T$ by comparing the first characters of adjacent suffixes.
$\setall$ denotes the set of all suffixes of $\T$, and also $\setL$, $\setS$, $\setLML$, and $\setLMS$ denote the set of all L-, S-, LML-, and LMS-suffixes of $\T$, respectively.
The size of a set $M$ is denoted by $\num{M}$.
Note that either $\num{\setL}$ or $\num{\setS}$ must be less than or equal to $N/2$ because all of the suffixes belong to either one.
Moreover, $\num{\setLMS}$ must be less than or equal to the smallest of $\num{\setL}$ and $\num{\setS}$.
For a subset $M$ of $\setall$ and a suffix  $\suf{j}$ of $M$, the rank of $\suf{j}$ is denoted by $\rank_M(j)$, namely, $\suf{j}$ is the $\rank_M(j)$-th smallest suffix of $M$.
When the context is clear, we denote $\rank_{\setall}$ as $\rank$.

For a subset $M$ of  $\setall$,
the suffix array $\SA_M$ of length $\num{M}$ is an integer array that stores
all pointers of $M$ such that corresponding suffixes are lexicographically sorted.
More precisely, for all suffixes $\suf{i}$ of $M$, $\SA_M[\rank_M(i)]=i$.
When the context is clear, we denote $\SA_{\setall}$ as $\SA$.
For each character $\vt$, the maximum interval in which the first characters of suffixes are equal to $\vt$ in $\SA$ is called the \textit{$\vt$-interval}.
Because L- and S-suffixes are respectively larger and smaller than their succeeding suffix, for any character $\vt$, L-suffixes that start with $\vt$ are always located before S-suffixes starting with $\vt$ in $\SA$.

The LCP array is an auxiliary array of $\SA$ such that $\LCP[i]$ contains the length of the longest common prefix of $\suf{\SA[i]}$ and $\suf{\SA[i-1]}$ for $1<i\leq N$, and $\LCP[1]=0$. See Figure~\ref{fig:example_of_SA}.

\section{Induced Sorting Framework}
\label{sec:previous}
Our algorithm is based on the induced sorting framework~\cite{KoA05,NongZC11,Nong13,FranceschiniM07}, so, in this section, we explain the algorithm in~\cite{NongZC11} as an example of the framework.
This algorithm runs in $O(N)$ time using $\sigma+N/\log N + O(1)$ extra words~\footnote{The space for storing beginning and ending positions of sub-arrays in recursive steps is not accounted for.}. 

The key point of the framework is to sort a subset of suffixes once and then sort another subset of suffixes from the sorted subset.
From this perspective, we say that the sorting of latter suffixes is induced from the former suffixes.
Let $\T^0$ be $\T$ and let $\T^{\ii+1}$ be a string such that $|\T^{\ii+1}|$ is the number of LMS-substrings of $\T^{\ii}$ and $\T^{\ii+1}[\jj]=k$, where the $j$-th LMS-substring from the left of $\T^{i}$ is the $k$-th lexicographically smallest LMS-substring of $\T^{\ii}$.
There are two important properties; the first is that $|\T^{\ii+1}| \leq \floor{|\T^{\ii}|/2}$ since the number of LMS-substrings in $\T^{\ii}$ is at most $\floor{|\T^{\ii}|/2}$, and the second is that the rank of the $j$-th LMS-suffix from the left of $\T^{\ii}$ within all LMS-suffixes in $\T^{\ii}$ corresponds to the rank of the $j$-th suffix from the left of $\T^{\ii+1}$ within all suffixes in $\T^{\ii+1}$.
The algorithm recursively computes the suffix array $\SA^i$ of the string $\T^i$ at each recursive step~$i$, namely, the algorithm sorts suffixes the number of which is smaller in more inner recursive steps.
Note that $\T^{\ii}$ has the same property of $\T$ such that $\T^{\ii}$ consists of an integer alphabet of $[1, \dots \sigma^\prime]$ for $1 \leq \sigma^\prime \leq |\T^{i}|$ and $\T^{\ii}$ contains $\sigma^\prime$  distinct characters.

Below is an overview of the algorithm for computing the $\SA^i$ of $\T^i$ at a recursive step~$i$.
Note that all suffixes and substrings that appear in the overview indicate those of $\T^i$.
\begin{enumerate}
	\item Sort all LMS-substrings. \label{step:sort_lms_sub}
	\item Sort all LMS-suffixes from sorted LMS-substrings. \label{step:sort_lms_suf}
	\item Sort all suffixes from sorted LMS-suffixes. \label{step:sort_suf}
	\begin{enumerate}
		\item \label{step:move_lms} Perform preprocessing for Step~\ref{step:sort_l}. 
		\item Sort all L-suffixes from sorted LMS-suffixes. \label{step:sort_l}
		\item Sort all S-suffixes from sorted L-suffixes. \label{step:sort_s}
	\end{enumerate}
\end{enumerate}
The essence of the algorithm is the part in which all suffixes are sorted from the sorted LMS-suffixes in Step~\ref{step:sort_suf}.
Step~\ref{step:sort_lms_sub} runs in almost the same way as Step~\ref{step:sort_suf}.
Step~\ref{step:sort_lms_suf} creates $\T^{\ii+1}$ and computes its suffix array recursively.
Therefore, we herein explain only Step~\ref{step:sort_suf}.
We describe Steps~\ref{step:sort_lms_sub}~and~\ref{step:sort_lms_suf} in Appendix~\ref{app:step12} to keep this paper self-contained.

We consider only the case of computing the $\SA^0=\SA$ of $\T^0=\T$ at the recursive step~0 since we can also compute the $\SA^\ii$ of $\T^\ii$ similarly at each recursive step $\ii$.
The algorithm requires three arrays, $\A$, $\lf/\rf$~\footnote{The notation was borrowed from ${\bf LF}/{\bf RF}$ used in~\cite{LiLH18}, which is the abbreviation of leftmost/rightmost free. Although the definition is the same as the $\mathit{bkt}$ array commonly used in previous research~\cite{NongZC11,Nong13,LouzaGT17}, the name ${\bf LF}/{\bf RF}$ is more specific. In our paper, we frequently use empty as the special symbol, so we prefer to use the notation $\lf/\rf$, which is the abbreviation for leftmost/rightmost empty.}, and $\type$.
$\A$ is an integer array of length $N$ to be $\SA$ at the end of the algorithm.
At the beginning of the algorithm, we assume that each $\A[i]$ is initialized to $\mathit{empty}$ in linear time, where empty is a special symbol that is used so that any element storing this symbol stores no meaningful value\footnote{Practically, the special symbol is represented as an integer $N+1$ indicating a position out of $\A$ so that we can distinguish the special symbol from pointers of $\A$.}.
$\type$ is a binary array of length $N$, which indicates the type of $\suf{j}$ such that $\type[j]=L$ if $\suf{j}$ is an L-suffix, and $\type[j]=S$ otherwise.
The $\type$ can be computed in $O(N)$ time with a right-to-left scan on $\T$ by comparing the first characters of the current suffix and its succeeding suffix.
$\lf/\rf$ is an integer array of length $\sigma$ such that $\lf[\vt]/\rf[\vt]$ indicates the next insertion position of a suffix starting with a character $\vt$ in $\A$.
$\lf[\vt]/\rf[\vt]$ is initially set to the head/tail of the $\vt$-interval of $\SA$, and it is managed in order to indicate the leftmost/rightmost empty position of the $\vt$-interval at any step of the algorithm.
$\lf$ can be initialized in $O(N)$ time as follows.
Let $\C_\vt$ be the number of suffixes starting with $\vt$.
First, $\lf[\vt]=\C_\vt$ is computed for all characters $\vt$ by counting $\headc{\ii}$ with a single scan on $\T$  by using $\lf[\headc{i}]$ as a counter, and last, $\lf[\vt]=1+\sum_{\vt^\prime < \vt} \C_{\vt^\prime}$ is computed by accumulating $\lf[\vt]=\C_\vt$ lexicographically.
Similarly, $\rf$ can also be computed in $O(N)$ time.

We assume that $\lf$ is initialized at the beginning of Step~\ref{step:sort_l} and that $\rf$ is also at the beginning of Step~\ref{step:move_lms} and Step~\ref{step:sort_s}.
During the steps, types of suffixes are obtained by $\type$.

\textbf{Step~\ref{step:move_lms}:}
As the result of Step~\ref{step:sort_lms_suf}, we have $\SA_{\setLMS}=\A[1 \dots \num{\setLMS}]$, and $\A[\num{\setLMS}+1 \dots N]$ is filled with empty.
With a right-to-left scan on $\SA_{\setLMS}$, we move each $\SA_{\setLMS}[i]=\suf{j}$ into $\A[\rf[\headc{j}]]$, which is the rightmost empty position of the $\headc{j}$-interval, and decrease $\rf[\headc{j}]$ by one to indicate the new rightmost empty position of the $\headc{j}$-interval.

\textbf{Step~\ref{step:sort_l}:}
With a left-to-right scan on $\A$, we read all L- and LMS-suffixes $\A[i]=\suf{j}$ in lexicographic order, if $\suf{j-1}$ is an L-suffix, store $\suf{j-1}$ in $\A[\lf[\headc{j-1}]]$, and increase $\lf[\headc{j-1}]$ by one.

\textbf{Step~\ref{step:sort_s}:}
This step runs almost the same way as Step~\ref{step:sort_l}.
With a right-to-left scan on $\A$, we read all L- and S-suffixes $\A[i]=\suf{j}$ in reverse lexicographic order, if $\suf{j-1}$ is an S-suffix, store $\suf{j-1}$ in $\A[\rf[\headc{j-1}]]$ and decrease $\rf[\headc{j-1}]$ by one.

Steps~\ref{step:move_lms}, \ref{step:sort_l}, and~\ref{step:sort_s} run in $O(N)$ time because each step scans $\A$ only one time, and any of the operations take constant time per access.
From Lemma~\ref{lem:induce_rank}, all induced L- and S-suffixes $\suf{\jj-1}$ are stored in $\A[\rank(\jj-1)]$, so $\A=\SA$ is obtained at the end of Step~\ref{step:sort_suf}.
Roughly speaking, the correctness of Lemma~\ref{lem:induce_rank} comes from the invariant that all suffixes stored in $\A$ are always sorted during the steps.
When reading $\A[\ii]=\suf{\jj}$, the L-suffix $\suf{\jj-1}$ must be larger than any suffix $\suf{\kk-1}$ already stored in $\headc{\jj-1}$-interval since $\suf{\kk}$ must appear at $\A[\ii^\prime]$ for $\ii^\prime < \ii$, and it holds that $\suf{\kk} < \suf{\jj}$ from the invariant.
Moreover, we do not miss any L-suffixes since we always store an induced L-suffix from a suffix stored in $\A[i]$ in a more rightward position $\A[\ii^\prime]$ for $\ii^\prime > \ii$.

\begin{lemma}[\cite{NongZC11}]
	\label{lem:induce_rank}
	When an L-suffix/S-suffix $\suf{\jj-1}$ is being induced in Step~\ref{step:sort_l}/Step~\ref{step:sort_s}, $\lf[\headc{\jj-1}]$/$\rf[\headc{\jj-1}]$ indicates $\rank(\jj-1)$.
\end{lemma}

Since $|\T^{i+1}| \leq \floor{|\T^i| / 2}$ and the algorithm runs in $O(|\T^i|)$ time for all $\ceil{\log N}$ recursive steps~$i$, the algorithm runs in $O(N)$ time in total.
Moreover, the algorithm requires $\sigma+N/\log N + O(1)$ extra words,
the first and second factors are for $\lf/\rf$ and $\type$, respectively.

\section{Optimal Time and Space Construction of Suffix Arrays}
\label{sec:new_algo}

We propose a novel algorithm for constructing suffix arrays on the basis of the induced sorting framework.
The space bottleneck of the previous algorithm~\cite{NongZC11} is the space of $\lf/\rf$ and $\type$.
Our algorithm embeds both arrays in the space of~$\A$, and runs in $O(N)$ time and in-place, namely, in optimal time and space.

As seen in Section~\ref{sec:previous}, the essence of the induced sorting framework is the part in which L-suffixes are sorted.
Therefore, we focus on how to sort the L-suffixes from sorted LMS-suffixes in $O(N)$ time and in-place.
We can also sort S-suffixes in the same way (see Appendix~\ref{sec:algo:sort_s}) and also LMS-substrings.
Thus, we have the following theorem.
\begin{theorem}
	Given a read-only string $\T$ of length $N$, which consists of integers $[1, \dots, \sigma]$ for $1 \leq \sigma \leq N$ and contains $\sigma$ distinct characters, there is an algorithm for computing the $\SA$ of $\T$ in $O(N)$ time and in-place.
\end{theorem}

\label{sec:algo:overview}
Our main idea for reducing the space is to store sorted L- and LMS-suffixes in three internal sub-arrays in $\A$.
We refer to these arrays as $\X$, $\Y$, and $\Z$.
The length of $\Y$ is $\sigma$, and for each character $\vt$, $\Y[\vt]$ stores either $\lf[\vt]$, the largest L-suffix starting with $\vt$, or the smallest LMS-suffix starting with $\vt$.
$\X$ and $\Z$ store all L- and LMS-suffixes other than the ones stored in $\Y$, respectively.

We embed $\lf$ in $\Y$.
Intuitively, this idea does not work because the total size of $\X$ and $\lf$ may exceed $\N$, and if so, $\X$ overlaps with $\lf$ in $\A$, and elements of $\lf$ required in the future may be overwritten by induced L-suffixes.
Moreover, we may not be able to even store  $\SA_{\setLMS}$ and $\lf$ in $\A$ at the same time before sorting L-suffixes because their total size can also be greater than $N$.
We avoid this problem by overwriting $\lf[\vt]$ only when it is no longer used in the future, namely, when  all L-suffixes starting with $\vt$ have been induced or there is no L-suffix starting with $\vt$.
We detect such timing by causing a conflict between induced L-suffixes.
Let $\CL_\vt$ be the number of L-suffixes starting with $\vt$.
We try to store all L-suffixes in $\X$, whose space is limited that can store only $\CL_\vt-1$ L-suffixes starting with $\vt$ for each character $\vt$.
More precisely, for a character $\vt$, the beginning and ending position of $\vt$-interval overlaps with the ending position of the preceding $\vt^\prime$-interval for $\vt^\prime < \vt$ and the beginning position of the succeeding $\vt^{\prime \prime}$-interval for $\vt^{\prime \prime} > \vt$, respectively.
Therefore, conflict must occur between the largest (the last induced) L-suffix starting with a character $\vt$ and the smallest (the first induced) L-suffix starting with $\vt^{\prime \prime} > \vt$.
We can detect the timing on the basis of conflicts, and we find that all L-suffixes starting with a smaller character $\vt$ have been induced and that $\lf[\vt]$ is no longer needed in the future.

We do not need $\type$ anymore for detecting the type of suffixes being induced.
We read L- and LMS-suffixes $\suf{\ii}$ stored either in $\X$, $\Y$, and $\Z$ in lexicographic order.
If $\suf{i}$ is read from $\X$ or $\Z$, we know the type of $\suf{i}$, so the type of $\suf{i-1}$ is easily obtained.
Otherwise, we do not know the type of $\suf{\ii}$ in $\Y$, so the type of $\suf{i-1}$ is non-trivial.
An important observation is that, for a suffix $\suf{i}$ in $\Y$, the preceding character $\headc{i-1}$ must be different from $\headc{i}$ since $\suf{i}$ is the largest L-suffix starting with $\headc{i}$ or the smallest LMS-suffix starting with $\headc{i}$.
Therefore, the type of $\suf{i-1}$ can be determined without $\type$ by comparing the characters $\headc{i-1}$ and $\headc{i}$.

We use $\typelf$ of $\sigma$ bits rather than $\N$ bits for distinguishing the L- and LMS-suffixes and the elements of $\lf$, which are stored in $\Y$.
Although $\typelf$ require $\sigma$ bits if it is stored naively, we can embed it in the space of $\Y$.
Details on the in-place implementation of $\typelf$ will be given in Section~\ref{sec:algo:typelf}.

\begin{figure}[tb]
  \centering
  \includegraphics[width=0.95\linewidth]{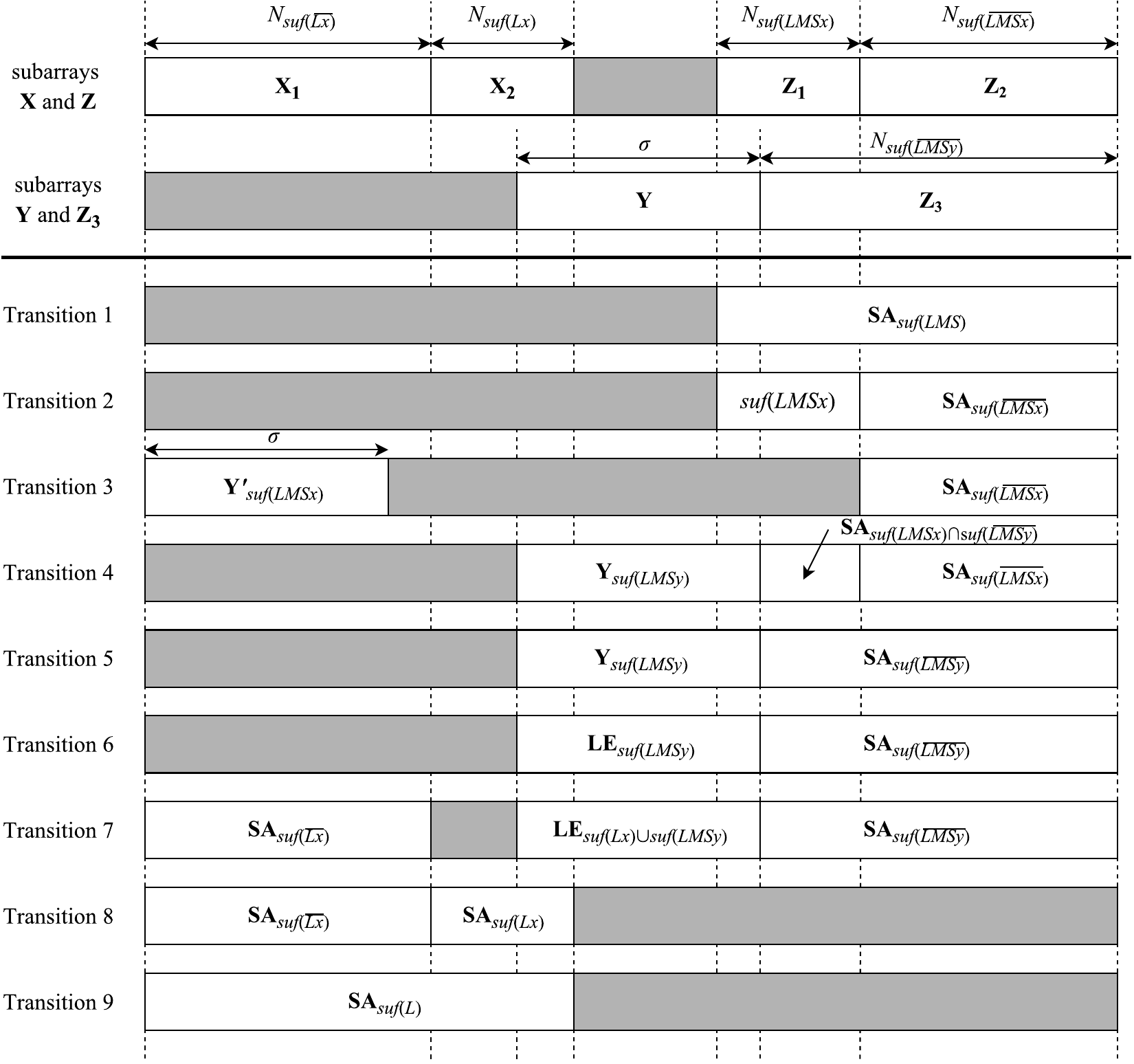}
  \caption{
    Inside transition of $\A$ while computing $\SA_{\setL}$ from $\SA_{\setLMS}$.
    Space colored with gray indicates empty space.
  }
  \label{fig:SALMS_to_SAL}
\end{figure}

The sub-arrays $\X$, $\Y$, and $\Z$ and their more internal sub-arrays are located in $\A$ as shown in Figure~\ref{fig:SALMS_to_SAL}.
The figure also shows all of the steps of our algorithm.
We partition the suffixes in certain subsets, which allows us to run each transition above in $O(N)$ time and in-place.
Let $\setxLMS$ be the set of the LMS-suffixes that are the smallest among all LMS-suffixes starting with the same character, and let $\setxLMSnot$ be the set of all other LMS-suffixes.
Let $\setyLMS$ be a subset of $\setxLMS$ that contains all $\suf{\ii} \in \setxLMS$ such that there is no L-suffix starting with $\headc{\ii}$, and let $\setyLMSnot$ be the set of all other LMS-suffixes.
We have $\setyLMS \subseteq \setxLMS$, $|\setxLMS| \le \sigma$, and $\setxLMSnot \subseteq \setyLMSnot$.
Let $\setxL$ be the set of all L-suffixes that are the largest among all L-suffixes starting with the same character, and let $\setxLnot$ be the set of all other L-suffixes.
Intuitively, $\setxL$ and $\setxLMS$ consist of L- and LMS-suffixes that are closest to the border of L- and S-suffixes in each $\vt$-interval in $\SA_{\setall}$, respectively.
Moreover, $\setxL \cup \setyLMS$ is made by selecting one suffix for each interval from the set $\setxL \cup \setxLMS$, where we give an L-suffix priority over an LMS-suffix if both exists.
Thus, we have $|\setxL \cup \setyLMS| \leq \sigma$.

We store various types of elements in $\Y$.
To reduce ambiguity, $\Y_M$ denotes $\Y$ \textit{overwritten} by a set of suffixes $M$ whose first characters are distinct (we consider that the initial $\Y$ is filled with empty).
More precisely, for a character $\vt$, $\Y_M[\vt]=\suf{i}$ if $\suf{i}$ starting with $\vt$ exists in $M$, and $\Y_M[\vt]=\Y[\vt]$ otherwise.
For example, $\Y_{\setyLMS}[\vt]=\suf{i}$ if $\suf{i} \in \setyLMS$ starting with $\vt$ exists, and $\Y_{\setyLMS}[\vt]$ is empty otherwise.
Moreover, $\lf_{\setyLMS}[\vt]=\suf{i}$ if $\suf{i} \in \setyLMS$ starting with $\vt$ exists, and $\lf_{\setyLMS}[\vt]=\lf[\vt]$ otherwise.

\subsection{Sort all L-suffixes}
\label{sec:algo:sort_l}

\newcommand{\dY}{\Y^\prime}

We compute $\SA_{\setL}$ in the head of $\A$ from $\SA_{\setLMS}$ stored in the head of $\A$, which is given by the result of sorting the LMS-suffixes.
The internal transitions of $\A$ in the algorithm are shown in Figure~\ref{fig:SALMS_to_SAL}, and each transition runs in $O(N)$ time and in-place.
In Transitions~\arabic{counterMakeSALMS}-\arabic{counterMakeLFLMS}, we compute $\lf$ and move LMS-suffixes in $\Y$ and $\Z$.
Transition~\arabic{counterMakeSALxnot} induces all L-suffixes from LMS-suffixes stored in $\Y$ and $\Z$ and stores them in $\X$ and $\Y$.
The concept of this transition is almost the same as Step~\ref{step:sort_l} in Section~\ref{sec:previous}.
In Transitions~\arabic{counterMakeSALx}-\arabic{counterMakeSAL}, we merge the L-suffixes of $\X$ and $\Y$ and obtain $\SA_{\setL}$.
The former part Transitions~\arabic{counterMakeSALMS}-\arabic{counterMakeSALMSxnot} is not so hard, so we omitted (we left the details in Appendix~\ref{sec:algo:sort_l_former}), and we only describe the latter part Transitions~\arabic{counterMakeLFLMS}-\arabic{counterMakeSAL} which is the most technical part of our algorithm.
We assume that we have a bit array $\typelf$ of $\sigma$ bits without extra space, and details on its in-place implementation are given in Section~\ref{sec:algo:typelf}.

As the result of Transitions~\arabic{counterMakeSALMS}-\arabic{counterMakeSALMSxnot}, we  have $\Y_{\setyLMS}$ for which $\Y[\vt]=\suf{i}$ if $\suf{i} \in \setyLMS$ starting with $\vt$ exists, and $\Y[\vt]$ is empty otherwise, and we also have $\typelf$ for which $\typelf[\vt]=1$ if an L-suffix starting with $\vt$ exists, and $\typelf[\vt]=0$ otherwise.

\makeLFLMSx
We transform $\Y_{\setyLMS}$ into $\lf_{\setyLMS}$.
With a right-to-left scan on $\T$, we compute $\CL_\vt$ for each character $\vt$ for which $\typelf[\vt]=1$, namely for which $\CL_\vt > 0$, and store it in $\Y[\vt]$ by using $\Y[\vt]$ as a counter.
Note that we never overwrite a suffix of $\setyLMS$ stored in $\Y[\vt]$ since  $\typelf[\vt]=0$ and there is no L-suffix starting with $\vt$.
With a left-to-right scan on $\Y$, we compute the prefix sum $\Y[\vt]=\lf[\vt]=1 + \sum_{\vt^\prime<\vt} \max (0, \CL_{\vt^\prime}-1)$ for each $\vt$ for which $\typelf[\vt]=1$.
Finally, we have  $\lf_{\setyLMS}$ in $\Y$ that  $\Y[\vt] = \lf[\vt]$ if $\typelf[\vt] = 1$, and $\Y[\vt]$ is $\suf{i} \in \setyLMS$ or empty otherwise.

\newcounter{cntAccess}
\newcounter{cntJudge}
\newcounter{cntStore}
\setcounter{tcounter}{1}
\setcounter{cntAccess}{\value{tcounter}}
\stepcounter{tcounter}
\setcounter{cntJudge}{\value{tcounter}}
\stepcounter{tcounter}
\setcounter{cntStore}{\value{tcounter}}
\makeSALxnot
We compute $\SA_{\setxLnot}$ in $\X_1$ and $\lf_{\setxL \cup \setyLMS}$ in $\Y$.
This transition consists of the following three parts whose concept is almost same as Step~\ref{step:sort_l} in Section~\ref{sec:previous}.
Part~\arabic{cntAccess} reads all L- and LMS-suffixes $\suf{i}$ lexicographically from $\X_1$, $\Y$, and $\Z_3$, Part~\arabic{cntJudge} judges whether $\suf{i-1}$ is an L-suffix or not, and Part~\arabic{cntStore} stores $\suf{i-1}$ if it is an L-suffix.
During the transition, we use $\typelf$ to determine whether $\Y[\vt]$ is a suffix (including both L- and LMS-suffixes) or an element of $\lf$.
The invariant is that $\typelf[\vt]=1$ if $\Y[\vt]$ is an element of $\lf$, and $\typelf[\vt]=0$ otherwise, that is, $\Y[\vt]$ is empty or a suffix of $\setxL \cup \setyLMS$.
As the result of the previous transition, $\typelf$ has already satisfied the invariant.

We explain Part~\arabic{cntAccess} last because it depends on Part~\arabic{cntStore}.

\textbf{Part~\arabic{cntJudge}: Judge whether $\suf{i-1}$ is an L-suffix or not.}
As already explained, the type of $\suf{\ii-1}$ can be obtained by comparing the first character $\headc{\ii-1}$ and its succeeding characters~$\headc{\ii}$.

\textbf{Part~\arabic{cntStore}: Store $\suf{i-1}$ if it is an L-suffix.}
We try to store an L-suffix $\suf{i-1}$ in $\X_1[\lf[\headc{i-1}]]$, which is the next insertion position of a suffix starting with $\headc{i-1}$.
If $\X_1[\lf[\headc{i-1}]]$ is empty, we simply store $\suf{i-1}$ there and increase $\lf[\headc{i-1}]$ by one  to update the next insertion position.
Otherwise, $\X_1[\lf[\headc{i-1}]]$ has already stored a suffix $\suf{\jj}$.
As already explained, a conflict must occur between the largest (the last induced) L-suffix starting with a character $\vt$ and the smallest (the first induced) L-suffix starting with $\vt^{\prime} > \vt$, and all L-suffixes starting with the smaller character $\vt$ have already induced.
Therefore, we compare the first characters $\headc{\ii-1}$ and $\headc{\jj}$, store the smaller one in $\lf[\min(\headc{\ii-1}, \headc{\jj})]$, store the larger one in $\X[\lf[\headc{\ii-1}]]$, and update $\typelf[\min(\headc{\ii-1}, \headc{\jj})]=0$.
Moreover, we increase $\lf[\headc{\ii-1}]$ by one if $\headc{\ii-1} > \headc{\jj}$.

\textbf{Part~\arabic{cntAccess}: Read all L- and LMS-suffixes $\suf{i}$ lexicographically.}
Recall that the arrays $\X_1$, $\Y$, and $\Z_3$ store sorted suffixes.
With a left-to-right scan on $\X_1$, $\Y$, and $\Z_3$, we scan $\X_1[i_X]$, $\Y[i_Y]$, and $\Z_3[i_Z]$ simultaneously in lexicographic order, where $i_X$, $i_Y$, and $i_Z$ are the scanning positions of $\X_1$, $\Y$, and $\Z_3$, respectively.
We recall the types of suffixes stored in $\X_1$, $\Y$, and $\Z_3$:
\begin{itemize}
	\item $\X_1[i_X]$ is either empty or an L-suffix.
	\item $\Y[i_Y]$ is either empty, a suffix of $\setxL \cup \setyLMS$, or $\lf[i_Y]$.
	\item $\Z_3[i_Z]$ is a suffix of $\in \setyLMSnot$.
\end{itemize}
Let $\vt_X$, $\vt_Y$, and $\vt_Z$ be the first characters of suffixes stored in $\X_1[i_X]$, $\Y[i_Y]$, and $\Z_3[i_Z]$, respectively, where $\vt_Y$ equals $i_Y$.
Here, we assume that $\vt_X$, $\vt_Y$, and $\vt_Z$ are $\sigma + 1 \not\in \Sigma$ if each index $i_X$, $i_Y$, or $i_Z$ indicates a position out of the corresponding array, that is, such $\vt_X$, $\vt_Y$, and $\vt_Z$ must not be chosen.
We also assume that $\vt_X$ is $\sigma + 1$ if $\X_1[i_X]$ is empty.

We choose the smallest character $\vt_i$ of $\vt_X$, $\vt_Y$, and $\vt_Z$. In case we need to break a tie, we give $\vt_X$ priority over $\vt_Y$, and $\vt_Y$ priority over $\vt_Z$.
Note that $\suf{\ii}$ is the smallest suffix of the three candidates because, for suffixes $\suf{j_1}$, $\suf{j_2}$, and $\suf{j_3}$ of $\setxLnot$, $\setxL \cup \setyLMS$, and $\setyLMSnot$, respectively, we have $\suf{j_1}<\suf{j_2} < \suf{j_3}$ if they all start with the same character, and $\suf{\ii}$ is chosen in this order of priority.
Next, we increase the scanning position by one.
Thus, we can read all L- and LMS-suffixes lexicographically.

One concern is that we may choose $\vt_Y=\ii_Y$ for which either $\Y[i_Y]$ is empty or $\Y[i_Y] = \lf[i_Y]$.
The former case implies that none of the L- or LMS-suffixes start with $\vt_Y$, so we increase $i_Y$ by one and choose the smallest character from the three candidates again.
In the latter case, let $\suf{\ii}$ be the largest L-suffix starting with $\vt$, it implies that $\typelf[i_Y]$ must be $1$, a conflict with $\suf{\ii}$ has not occured yet, and $\suf{\ii}$ has already been read and is still stored in $\X_1$.
So, in this case also, we increase $i_Y$ by one and choose the smallest character from the three candidates again.
$\suf{\ii}$ stored in $\X_1$ will conflict with another L-suffix and be stored in $\lf[\vt_Y]$ in the future.

\makeSALx
We compute $\SA_{\setxL}$ in $\X_2$ and initialize the space except for $\X$ in $\A$ as empty.
All L-suffixes of $\setxL$ are stored in $\Y$, and we have $\typelf$ for which $\typelf[\vt]=1$ if $\Y[\vt]$ stores an L-suffix of $\setxL$, and $\typelf[\vt]=0$ otherwise.
With a left-to-right scan on $\Y$, we move all L-suffixes $\suf{\ii}$ for which $\typelf[\headc{\ii}]=1$ in back of $\X_1$ while preserving the order, and we obtain $\SA_{\setxL}$ in $\X_2$.
Finally, we fill $\A[\num{\setL} \dots N]$ with empty.

\makeSAL
We compute $\SA_{\setL}$.
By applying the in-place stable merge algorithm in Theorem~\ref{thm:merge_inplace} to $\SA_{\setxL}$ and $\SA_{\setxLnot}$ considering the first characters as keys, we compute $\SA_{\setL}$ in $O(N)$ time and in-place.

\begin{theorem}[\cite{Chen03}]
  \label{thm:merge_inplace}
  For two sorted integer arrays $\A_1=\A[1\dots N_1]$ and $\A_2=\A[N_1+1 \dots N_1+N_2]$ that are stored in an array $\A[1 \dots N_1 + N_2]$, there is an in-place linear time ($O(\N_1 + \N_2)$ time) algorithm that can stably merge $\A_1$ and $\A_2$ in $\A$.
\end{theorem}

\textbf{Remark:}
For ease of explanation, we use the complex stable merge algorithm in Transition~\arabic{counterMakeSALMSxnot} and~\arabic{counterMakeSAL} for sorting L-suffixes.
We can optimize the algorithm so that the algorithm does not use the merge algorithm for sorting L-suffixes and use only two times for sorting S-suffixes.
Since $\SA_{\setyLMSnot}$ is read only sequentially, we can simulate the sequential scan of $\SA_{\setyLMSnot}$ by scanning $\SA_{\setxLMS \cap \setyLMSnot}$ and $\SA_{\setxLMSnot}$ sequentially.
Moreover, Transition~\arabic{counterMakeSAL} is equal to Transition 2 in sorting S-suffixes (see Appendix~\ref{sec:algo:sort_s}), so we can skip Transition~\arabic{counterMakeSAL} and avoid to use the stable merge algorithm.

\subsection{In-place Implementation of $\typelf$}
\label{sec:algo:typelf}
We store suffixes and elements of $\lf$ in $\Y$ in a compact representation so that whose most significant bits (MSBs) are vacant, and embed $\typelf$ in the MSBs of $\Y$.
Since each original value can be obtained from the simple compact representation in $O(1)$ time, it does not cause any problems for all transitions shown in Figure~\ref{fig:SALMS_to_SAL}.

$\lf$ is a non-decreasing sequence, so we remember the leftmost $\mm$-interval that includes the position $2^{\ceil{\log \N}-1}$ in $\X_1$ whose MSB is $1$, and also remember the MSB of $\lf[\mm]$ as $\mathit{msb}$.
In Transition~\arabic{counterMakeSALxnot}, $\mathit{msb}$ is initially $0$ but finally becomes $1$.
All elements of $\lf[\vt]$ are stored in $\Y$ in the compact representation by clearing the MSBs to $0$.
The original value of each $\lf[\vt]$ can be obtained in $O(1)$ time as follows;
\begin{itemize}
  \item Set the MSB to $0$ for $\vt < \mm$.
  \item Set the MSB to $\mathit{msb}$ for $\vt=\mm$.
  \item Set the MSB to $1$ for $\vt > \mm$.
\end{itemize}
A suffix $\suf{\ii}$ is stored as $\floor{\ii/2}$ so that the MSB is vacant.
We use two important properties to obtain original values that, for a suffix $\suf{\ii}$ stored in $\Y[\vt]$, (1) the first character of $\suf{\ii}$ must be $\vt$, and (2) the preceding character $\headc{\ii-1}$ does not equal $\vt$ (since $\suf{\ii}$ is the largest L-suffix starting with $\vt$ or the smallest LMS-suffix starting with $\vt$).
We can obtain an original suffix $\suf{\ii}$ from its compact representation $\Y[\vt]=\jj$.
The candidate of $\ii$ is $2\jj$ or $2\jj + 1$.
If $\headc{2\jj} \neq \headc{2\jj+1}$, we choose one that equals $\vt$ with Property~1.
Otherwise, we choose $2\jj$ with Property~2.

Thus, we can store all elements of $\lf$ and suffixes in $\Y$ in a compact representation whose MSBs are vacant and store $\typelf$ in-place in the MSBs of $\Y$.

\section{Optimal Time and Space Construction of Suffix Arrays and LCP Arrays}
\label{sec:lcp}
We propose an algorithm for computing the suffix array and LCP array of a given read-only string $\T$ in $O(N)$ time and in-place.
We revisit Manzini's algorithm~\cite{Manzini04}, which constructs an LCP array $\LCP$ from a given string $\T$ and a suffix array $\SA$ in $O(N)$ time by using $\sigma + O(1)$ extra words.
The algorithm uses a $\psi$ array $\psia$ which is also called the \textit{rank next array}, where  $\psia[\rank(i)]=\rank(i+1)$ for $1 \leq i < N$.
The algorithm consists of two parts.
The first part computes $\psia$ in $O(N)$ time by using $\sigma+O(1)$ extra words.
The second part converts $\psia$ into $\LCP$ in $O(N)$ time and in-place.
Therefore, $\LCP$ can be computed in $O(N)$ time and in-place if $\psia$ can be computed in $O(N)$ time and in-place.

\newcommand{\isa}{{\bf ISA}}

Let $\A$ and $\B$ be integer arrays of length $N$ to be $\SA$ and $\LCP$ at the end of the algorithm, respectively.
Our algorithm computes $\B=\psia$ with both arrays $\A$ and $\B$ and $O(1)$ extra words.
After that, it computes $\A=\SA$ in-place as described in Section~\ref{sec:new_algo} and converts $\B=\psia$ into $\B=\LCP$ in-place as in Manzini's way.
For computing $\psia$, we use the inverse suffix array $\isa$ such that $\isa[\SA[i]]=i$, which is also called the \textit{rank array} since $\isa[i]=\rank(i)$.
The algorithm runs in the following steps.
\begin{enumerate}
  \item Compute $\B=\SA$.
  \item Compute $\A=\isa$ from $\SA$.
  \item Compute $\B=\psia$, that is, drop $\SA$.
  With a left-to-right scan on $\isa$, set $\B[\isa[i]]=\isa[i+1]$ if $\isa[i]<N$.
  \item Compute $\A=\SA$ as described in Section~\ref{sec:new_algo}.
  \item Convert $\B=\psia$ into $\B=\LCP$ as in Manzini's way.
\end{enumerate}
All of the steps run in $O(N)$ time and in-place.
Thus, we have the following theorem.
\begin{theorem}
  Given a read-only string $\T$ of length $\N$, which consists of integers $[1, \dots, \sigma]$ for $1 \leq \sigma \leq \N$ and contains $\sigma$ distinct characters, there is an algorithm for computing both $\SA$ and $\LCP$ of $\T$ in $O(\N)$ time and in-place.
\end{theorem}

\bibliography{ref}
\appendix
\section{Appendix}
\label{sec:appendix}

\subsection{Details on Induced Sorting Framework}
\label{app:step12}
We describe Steps~\ref{step:sort_lms_sub}~and~\ref{step:sort_lms_suf} in the induced sorting framework.
A more detailed overview of the algorithm~\cite{NongZC11} is given as follows.

\begin{enumerate}
	\item Sort all LMS-substrings. 
	\begin{enumerate}
		\item Sort all LMS-substrings by their first characters. \label{step:sort_lms_head}
		\item Sort all L-substrings from LMS-substrings sorted by their first characters. \label{step:sort_l_sub}
		\item Sort all S-substrings from sorted L-substrings. \label{step:sort_s_sub}
	\end{enumerate}

	\item Sort all LMS-suffixes from sorted LMS-substrings.
	\begin{enumerate}
		\item Check whether or not all LMS-substrings are unique, and if so, do nothing and go to Step~\ref{step:sort_suf}.
		\item Create a new string $\T^{i+1}$.
		\item Compute $\SA^{i+1}$ of $\T^{i+1}$ recursively. \label{step:recursion}
		\item Convert $\SA^{i+1}$ into the suffix array of LMS-suffixes. \label{step:create_salms}
	\end{enumerate}

	\item Sort all suffixes from sorted LMS-suffixes. 
	\begin{enumerate}
		\item Perform preprocessing for Step~\ref{step:sort_l}. 
		\item Sort all L-suffixes from sorted LMS-suffixes. 
		\item Sort all S-suffixes from sorted L-suffixes.
	\end{enumerate}
\end{enumerate}

\begin{figure}[tb]
	\centering
	\includegraphics[width=0.9\linewidth]{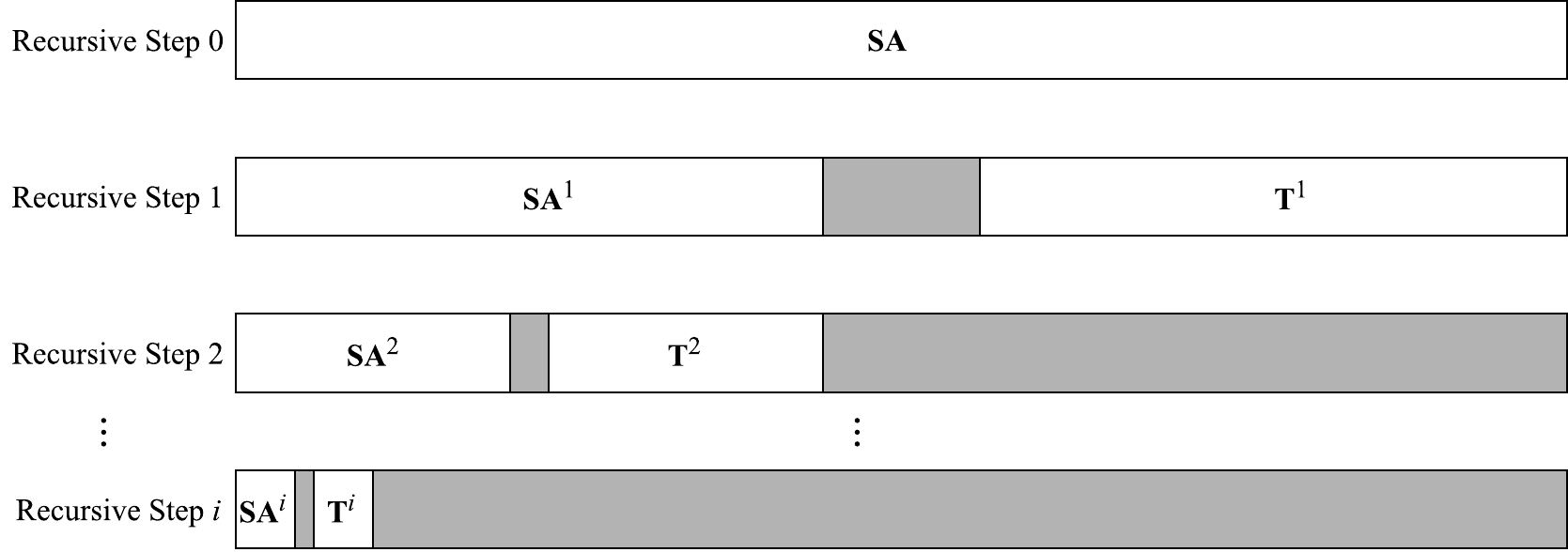}
	\caption{
		Inside transition of $\A$ at each recursive step $i$.
		Space of $\SA^i$ at recursive step $i$ is used to store $\SA^{i+1}$ and $\T^{i+1}$ for next recursive step $i+1$.
	}
	\label{fig:induced_sort_rec}
\end{figure}

We assume that $\lf$ is initialized at the beginning of Step~\ref{step:sort_l_sub} and $\rf$ as well at the beginning of Steps~\ref{step:sort_lms_head} and~\ref{step:sort_s_sub}.

\textbf{Step~\ref{step:sort_lms_head}:}
With a right-to-left scan on $\T$, we store all LMS-substrings $\sub{i}$ in $\A[\rf[\headc{i}]]$ and  decrease $\rf[\headc{i}]$ by one.

\textbf{Steps~\ref{step:sort_l_sub} and~\ref{step:sort_s_sub}:}
Both are almost the same as Steps~\ref{step:sort_l} and~\ref{step:sort_s}, respectively.
The important difference is that all substrings other than LMS-substrings are deleted at the end of Step~\ref{step:sort_lms_sub}.
In Step~\ref{step:sort_l_sub}, after inserting an L-substring $\sub{\jj-1}$ induced from an L- or LMS-substring $\A[i]=\sub{j}$ into $\A[\lf[\headc{\jj-1}]]$, $\A[i]$ is set to empty.
After that, $\A$ contains only sorted LML-substrings.
In Step~\ref{step:sort_s_sub}, after inserting an S-substring $\sub{j-1}$ induced from an LML- or S-substring $\A[i]=\sub{j}$ into $\A[\rf[\headc{j-1}]]$, $\A[i]$ is set to empty.
At the end of Step~\ref{step:sort_lms_sub}, $\A$ contains only sorted LMS-substrings.

\textbf{Step~\ref{step:sort_lms_suf}:}
If LMS-substrings are all unique, pointers of sorted LMS-substrings are equal to those of sorted LMS-suffixes, so we go to Step~\ref{step:sort_suf} immediately.
Otherwise, we create $\T^1$ such that each LMS-substring in $\T$ is replaced by its rank among LMS-substrings and compute the suffix array $\SA^1$ of $\T^1$ recursively.
$\SA^1$ and $\T^1$ are stored in $\A[1 \dots |\T^1|]$ and $\A[N-|\T^1|+1 \dots N]$, respectively, and they never overlap because $|\SA^1|=|\T^1|=\num{\setLMS}$ is less than or equal to $N/2$.
See Figure~\ref{fig:induced_sort_rec}.
Here, a pointer of $\T^1[i]$ is related to the pointer of the $i$-th LMS-substring from the left in $\T$.
After $\SA^1$ is computed, we convert all pointers of $\SA^1$ into the corresponding ones of $\SA_{\setLMS}$.

\subsection{Sort all L-suffixes: Former Transitions}
\label{sec:algo:sort_l_former}
We describe Transitions~\arabic{counterMakeSALMS}-\arabic{counterMakeSALMSxnot}, which are omitted in Section~\ref{sec:algo:sort_l}.

\makeSALMS
We shift $\SA_{\setLMS}$ stored in the head of $\A$ into $\Z$.

\makeSALMSynot
We store $\setxLMS$ in $\Z_1$ and compute $\SA_{\setxLMSnot}$ in $\Z_2$.
Note that the suffixes in $\Z_2$ are sorted but may not be in $\Z_1$.
Let $j$ be the insertion position in $\Z_2$ for $\SA_{\setxLMSnot}$, which is initially set to $\num{\setLMS}$, namely, the end of $\Z$.
With a right-to-left scan on $\Z=\SA_{\setLMS}$, we swap $\SA_{\setLMS}[i]=\suf{\kk}$ with $\Z[j]$ and decrease $j$ by one if $\suf{\kk} \in \setxLMSnot$ and do nothing otherwise.
Whether or not $\suf{\kk}$ belongs to $\setxLMSnot$ can be judged in $O(1)$ time by comparing the first characters because the first characters $\headc{\SA_{\setLMS}[i]}$ and $\headc{\SA_{\setLMS}[i-1]}$ are the same if and only if $\suf{\kk}\in \setxLMSnot$.
Since we shift the suffixes of $\setxLMSnot$ to the end of $\Z$ while preserving the order of the shifted suffixes, we obtain $\setxLMS$ in $\Z_1$ (which may not be sorted) and $\SA_{\setxLMSnot}$ in $\Z_2$.

Unfortunately, we cannot compute $\Y_{\setyLMS}$ at this point directly because we currently do not know the size of $\num{\setyLMSnot}$
determining the starting position of $\Y$ within $\A$.
We obtain this information in Transition~\arabic{counterMakeYLMSx}.
To start with, we consider a temporary array $\dY=\A[1 \dots \sigma]$ and compute $\dY_{\setxLMS}$.

\makeYLMSy
We compute $\dY_{\setxLMS}$.
With a right-to-left scan on $\Z_1$, we try to move $\Z_1[i]=\suf{j_1}$ into $\dY[\headc{j_1}]$.
However, $\dY[\headc{j_1}]$ may contain an LMS-suffix $\suf{j_2}$ because $\dY$ may overlap with $\Z_1$.
We simply move $\suf{j_1}$ into $\dY[\headc{j_1}]$ if $\dY[\headc{j_1}]$ is empty and do nothing if $\dY[\headc{j_1}]$ is $\suf{j_1}$ because then $\Z_1[i]$ and $\dY[\headc{j_1}]$ are the same entry in $\A$.
Otherwise, $\dY[\headc{j_1}]$ contains a suffix $\suf{j_2}$ such that $\suf{j_2} \neq \suf{j_1}$.
In this case, we move $\suf{j_1}$ into $\dY[\headc{j_1}]$ and then try to move $\suf{j_2}$ into $\dY[\headc{j_2}]$.
We repeat this procedure until we move $\suf{j_k}$ to  $\dY[\headc{j_k}]$, which is empty, or encounter $\dY[\headc{j_k}]=\suf{j_k}$.
Because $\num{\setxLMS} \leq \sigma$ and the first characters of $\setxLMS$ are all different, the number of insertions is $O(\sigma)$, and this transition can be done in $O(\sigma)$ time.
Finally, we have $\dY_{\setxLMS}$ such that $\dY[\headc{i}] = \suf{i}$ if $\suf{i} \in \setxLMS$ or $\dY[\headc{i}]$ is empty otherwise.

\makeYLMSx
We compute $\Y_{\setyLMS}$ and $\SA_{\setxLMS \cap \setyLMSnot}$.
The set $\setxLMS \cap \setyLMSnot$ consists of each $\setxLMS$ suffix for which there is an L-suffix starting with the same character, and $\setyLMS$ is other $\setxLMS$.
We compute $\typelf[\vt]=1$ if there is an L-suffix starting with $\vt$, and $\typelf[\vt]=0$ otherwise.
We initialize $\typelf$ with $0$.
With a right-to-left scan on $\T$, we set $\typelf[\vt]=1$ for an L-suffix starting with $\vt$.
Now we know that a suffix stored in $\dY_{\setxLMS}[\vt]$ with $\typelf[\vt]=1$ belongs to $\setyLMSnot \cap \setxLMS$.
With a right-to-left scan on $\dY_{\setxLMS}$, we move such suffixes in front of $\Z_2=\SA_{\setxLMSnot}$ while preserving the order; then, we have $\Z_1=\SA_{\setxLMS \cap \setyLMSnot}$.
We just move $\dY$ in front of $\Z_1$, and we have $\Y_{\setyLMS}$.

\makeSALMSxnot
We compute $\SA_{\setyLMSnot}$.
Because a suffix $\suf{\ii}$ of $\setyLMSnot \cap \setxLMS$ is smaller than all suffixes of $\setxLMSnot$ starting with the same character $\headc{\ii}$, $\SA_{\setyLMSnot}$ can be obtained by stably merging the last two arrays $\SA_{\setyLMSnot \cap \setxLMS}$ and $\SA_{\setxLMSnot}$ with respect to the first characters as keys.
The merged array contains \emph{all} $\setyLMSnot$ suffixes since $(\setyLMSnot \cap \setxLMS) \cup \setxLMSnot = \setyLMSnot$.
By applying Theorem~\ref{thm:merge_inplace} to $\SA_{\setyLMSnot \cap \setxLMS}$ and $\SA_{\setxLMSnot}$, we compute $\SA_{\setyLMSnot}$ in $O(N)$ time and in-place.

\subsection{Sort all S-suffixes}
\label{sec:algo:sort_s}

\begin{figure}[tb]
  \centering
  \includegraphics[width=0.95\linewidth]{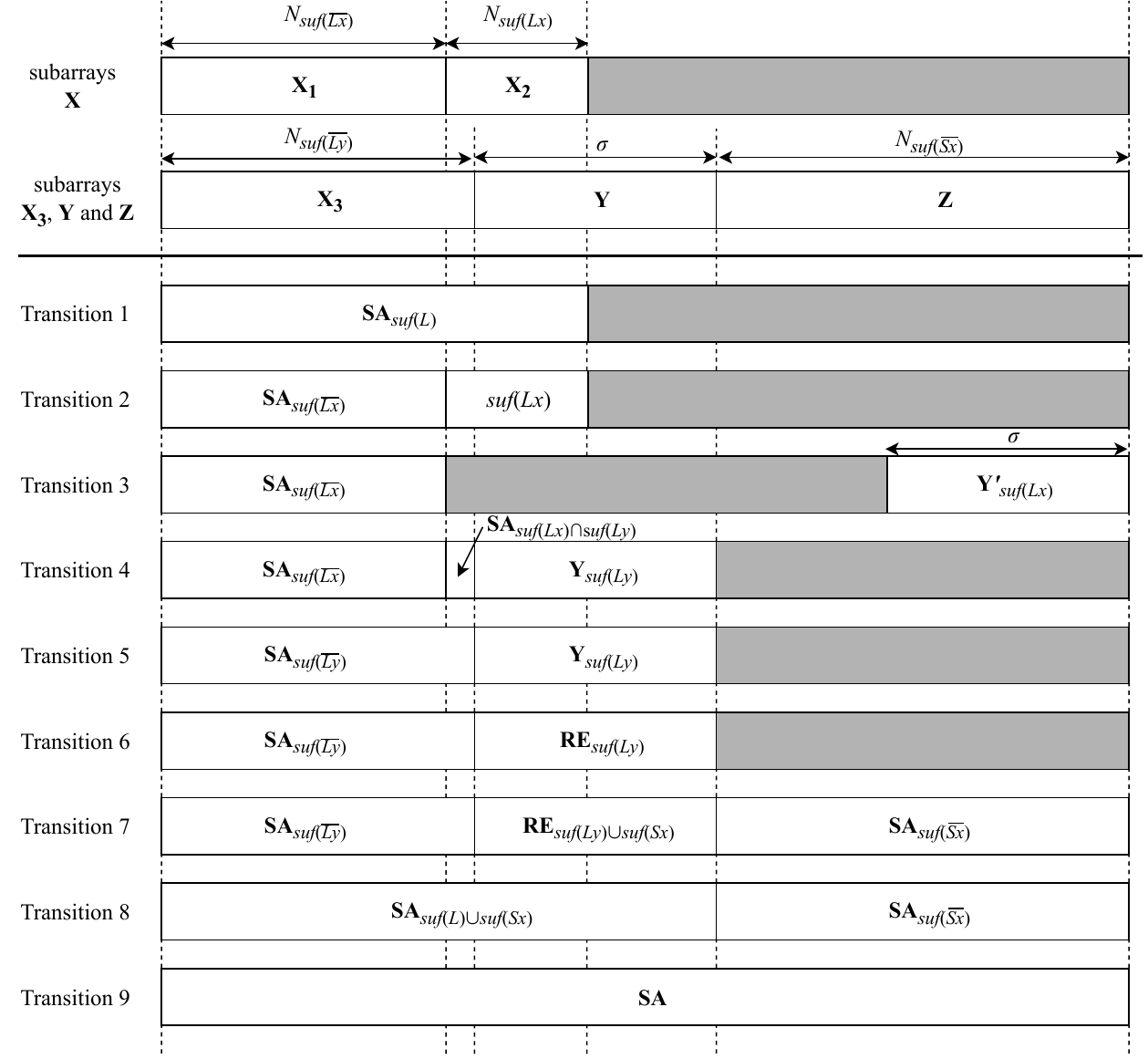}
  \caption{
    Inside transition of $\A$ while computing $\SA$ from $\SA_{\setL}$.
    Space colored with gray indicates empty space.
  }
  \label{fig:SAL_to_SA}
\end{figure}

We can sort all S-suffixes in almost the same way as sorting L-suffixes but compute $\SA$ instead of $\SA_{\setS}$.
The same can be said by switching the roles of $\lf$, L- and LMS-suffixes with $\rf$, S-, and L-suffixes, respectively.
Let $\setxS$ be the smallest suffixes starting with each character $\vt$, let $\setxSnot$ be the set of the other S-suffixes, let $\setyL$ be the set of the largest L-suffixes starting with each character $\vt$ such that no S-suffix starts with $\vt$, and let $\setyLnot$ be the set of the other L-suffixes.
We compute $\SA_{\setyLnot}$, $\rf_{\setyL \cup \setxS}$, and $\SA_{\setxSnot}$ from $\SA_{\setL}$ in a similar way as Transitions~\arabic{counterMakeSALMS}-\arabic{counterMakeSALxnot} in Section~\ref{sec:algo:sort_l}.
Note that $\rf_{\setyL \cup \setxS}$ equals $\SA_{\setyL \cup \setxS}$ from the definition.
We compute $\SA$  in $O(N)$ time and in-place by considering the first characters as keys, by applying Theorem~\ref{thm:merge_inplace} to $\SA_{\setyLnot}$ and $\SA_{\setyL \cup \setxS}$, and then by applying the result and  $\SA_{\setxSnot}$.

Thus, all S-suffixes can be sorted in $O(N)$ time and in-place as in Section~\ref{sec:algo:sort_l}.
See Figure~\ref{fig:SAL_to_SA}.

\subsection{Store $O(\log N)$ Integers for Recursion}
\label{sec:recursion}
We propose a simple technique for storing the locations of $\T^i$ and $\SA^i$ of $\A$ and getting the values in $O(1)$ time and in-place.

Let $N_i$ be  $|\T^i|$.
The key property is that $N_{i+1}$ is at most $\floor{N_i/2}$ because $N_{i+1}$ equals the number of LMS-substrings of $\T^i$.
We store $\SA^i$ and $\T^i$ in $\A[1 \dots N_i]$ and $\A[M_i + 1 \dots M_i + N_i]$, where $M_i = \floor{N / 2^i}$ is the upper bound of $N_i$.
Each $M_i$, which is the beginning position of $\T^i$ in $\A$, can be computed in $O(1)$ time and in-place by right-shifting $N$ by $i$ bits .

Let $\full$ be a binary array such that $\full[i]=1$ if $N_i=M_i$ and 0 otherwise.
If $\full[i]=1$, we do not need to store $N_i$ because $N_i=M_i$ and $M_i$ can be obtained in $O(1)$ time and in-place.
Otherwise, $\A[N_i+1 \dots M_i]$ is unused for computing the $\SA^\ii$ of $\T^\ii$, so we store $N_i$ in $\A[M_i]$.
Because the number of recursions is at most $\ceil{\log N}$, $\full$ can be stored in a word.
Thus, we can store and get each $N_i$, which is the length of $\T^i$, in $O(1)$ time and in-place.

\end{document}